\documentclass[aps,prl,twocolumn,superscriptaddress,amsmath,amssymb]{revtex4}

\usepackage{graphicx}
\usepackage{dcolumn}
\usepackage{bm}
\usepackage{amsmath}
\usepackage{amssymb}
\usepackage{epsfig}
\usepackage{epstopdf} 
\usepackage{array}
\usepackage{subfigure}

\begin{document} 
\title{Optical study of archetypical valence-fluctuating Eu-systems} 

\author{V. Guritanu} 
\author{S. Seiro}
\author{J. Sichelschmidt}
\author{N. Caroca-Canales}
\affiliation{Max Planck Institute for Chemical Physics of Solids, 01187 Dresden, Germany} 
\author{T. Iizuka}
\author{S. Kimura}
\affiliation{UVSOR Facility, Institute for Molecular Science, Okazaki 444-8585, Japan}
\author{C. Geibel}
\author{F. Steglich}
\affiliation{Max Planck Institute for Chemical Physics of Solids, 01187 Dresden, Germany}

\date{\today }\texttt{}

\begin{abstract}
We have investigated the optical conductivity of the prominent valence fluctuating compounds EuIr$_2$Si$_2$ and EuNi$_2$P$_2$ in the infrared energy range to get new insights into the electronic properties of valence fluctuating systems. For both compounds we observe upon cooling the formation of a renormalized Drude response, a partial suppression of the optical conductivity below 100\,meV and the appearance of a mid-infrared peak at 0.15\,eV for EuIr$_2$Si$_2$ and at 0.13\,eV for EuNi$_2$P$_2$.  Most remarkably, our results show a strong similarity with the optical spectra reported for many Ce- or Yb-based heavy fermion metals and intermediate valence systems, although the phase diagrams and the temperature dependence of the valence differ strongly between Eu- and Ce-/Yb-systems. This suggests that the hybridization between 4$f$- and conduction electrons, which is responsible for the properties of Ce- and Yb-systems, plays an important role in valence fluctuating Eu-systems. 
 \end{abstract}

\maketitle

\begin{figure}[bht]
\centerline{\includegraphics[width=8cm,clip=true]{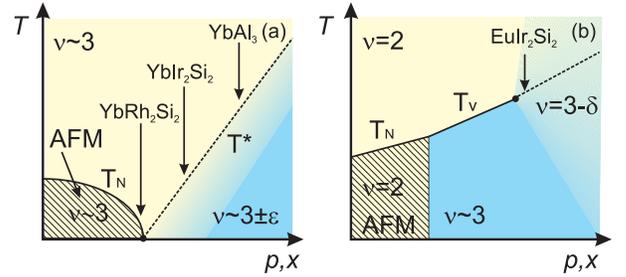}}
 \caption{(Color online) 
Schematic phase diagrams of (a) KL/IV Ce-($\nu$\,$\sim$\,3\,-\,$\epsilon$)
 or Yb-($\nu$\,$\sim$\,3\,+\,$\epsilon$) based systems and (b) VF Eu-based systems; [$\epsilon, \delta$\,=\,$f(x,p,T)$]. 
} 
\label{phase} 
\end{figure}

Intermetallic compounds based on rare earth elements with an unstable valence, especially Ce, Eu and Yb, present many unusual properties and are therefore the subject of intense research since many years. Historically, the understanding of these compounds has been dominated by a dichotomy between the Kondo lattice/intermediate valence (KL/IV) scenario for  Ce-/Yb-based systems and the valence fluctuating (VF) scenario for Eu-based systems. In essence, the former corresponds to a quantum mixing of two valence states induced by the hybridization $V_{fc}$ between localized $f$ electrons and delocalized conduction electrons, while the latter corresponds to thermal fluctuation between two integer-valence states. The best criterion to differentiate between these two cases is the evolution of the valence $\nu$ as a function of temperature $T$ and composition $x$ or pressure $p$. In KL/IV systems increasing the hybridization leads to a smooth and continuous evolution of $\nu$ as a function of $p$ or $x$. As a result the magnetic order observed at small $V_{fc}$, i.e. for an almost stable trivalent Ce or Yb state, disappears smoothly with the ordering temperature $T_{\rm N}$ continuously decreasing to $T$\,=\,0 at a quantum critical point, where huge correlations effects occur [Fig. \ref {phase}(a)] \cite{si2010}. Even beyond the quantum critical point the change of $\nu$ as a function of temperature is small, usually less than 0.1 between 0 and 300\,K. In contrast, VF Eu-systems present a pronounced first-order valence transition as a function of $p$ or $x$ and temperature from a divalent Eu-state with a large-moment magnetic order to an almost trivalent state showing Van-Vleck paramagnetism [Fig. \ref {phase}(b)] \cite{segre1982, seiro2011}. The strong first-order transition suppresses fluctuations and therefore correlation effects are rather weak, even close to the critical end point of the first-order phase boundary. Well beyond this critical end point in Eu-systems the change of $\nu$ is typically still larger than 0.5 between 0 and 300\,K. 

For Ce- or Yb-based KL/IV systems the nature of the electronic states and the low-energy excitations have been intensively investigated and are relatively well understood (with the exception of the immediate vicinity of the quantum critical point) while for Eu-based VF systems this knowledge is very limited. In particular, the optical conductivity has been studied in great detail in many Ce- and Yb-based KL/IV compounds \cite{garner2000,mena2005,okamura2004,degiorgi2001}, but no such studies have yet been published for Eu-based VF systems. 

In this Letter, we present a study of the electrodynamic response of two prominent VF Eu-systems, EuIr$_2$Si$_2$ and EuNi$_2$P$_2$. Our results reveal  an optical conductivity that is surprisingly similar to that of KL/IV systems: upon decreasing temperature we observe in both Eu-compounds the formation of a renormalized Drude response,  a partial suppression of the optical conductivity and a well-defined mid-infrared (MIR) peak at about 150\,meV. This is in strong contrast to the huge difference in the temperature dependence of the valence between Ce- or Yb-based KL/IV systems and Eu-based VF systems.

\begin{figure}[bht]
\centerline{\includegraphics[width=10cm,clip=true]{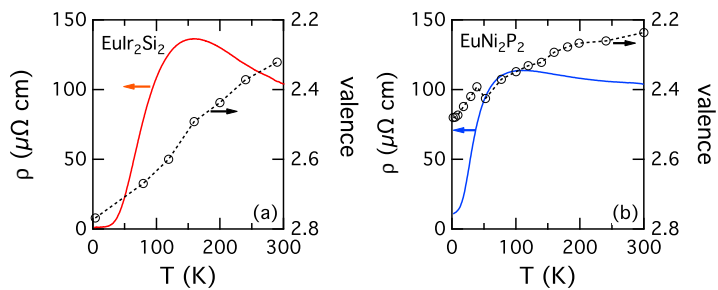}}
 \caption{(Color online) 
Resistivity and valence as a function of temperature of  EuIr$_2$Si$_2$ and EuNi$_2$P$_2$. The valence data are taken from Refs.\cite{chevalier1986,nagarajan1985}.}
\label{rho_valence} 
\end{figure}

EuIr$_2$Si$_2$ is an archetypical VF Eu-system with a valence deduced from M\"{o}ssbauer experiments \cite{chevalier1986} that decreases from 2.8 at 4\,K to 2.3 at 300\,K [Fig. \ref {rho_valence}(a)]. EuNi$_2$P$_2$ presents a unique situation among Eu-based VF systems. At high temperatures ($T$\,$>$\,100\,K) it is similar to EuIr$_2$Si$_2$, but at low temperature its valence \cite{nagarajan1985}, instead of approaching 3, remains close to 2.5, see Fig. \ref {rho_valence}(b). The reason why at low temperatures EuNi$_2$P$_2$ differs from standard VF Eu-systems like EuCu$_2$Si$_2$, EuNi$_2$Si$_2$ or EuIr$_2$Si$_2$, is presently not clear. As shown in Fig. \ref {rho_valence}, for both EuIr$_2$Si$_2$ and EuNi$_2$P$_2$ compounds the resistivity $\rho(T)$ increases upon cooling below 300 K and passes through a broad maximum at about 150\,K (100\,K) in EuIr$_2$Si$_2$ (EuNi$_2$P$_2$) before decreasing dramatically towards low temperatures. This temperature dependence reflects the valence fluctuations \cite{zipper1987}, but also looks quite similar to that observed for Yb-/Ce-based IV systems.  

Single crystals of EuIr$_2$Si$_2$ and EuNi$_2$P$_2$ were synthesized as described in Refs. \cite{seiro2011, danzenbacher2009}. Near-normal incidence reflectivity spectra were measured using a Fourier transform spectrometer Bruker IFS 66 v/S at energies 10\,meV-0.8\,eV in the temperature range from 6 to 260\,K. To obtain the absolute value of the reflectivity a gold layer was evaporated $\it{in}$ $\it{situ}$ on the crystal surface. At room temperature the reflectivity was measured for energies 0.6-1.25\,eV using a JASCO FTIR 610 spectrometer, where an Al mirror was used as a reference, and from 1.2\,eV to 30\,eV using synchrotron radiation at the beamline 7B of UVSOR-II, Institute for Molecular Science. For EuIr$_2$Si$_2$ the reflectivity measurements were extended down to 4\,meV using a JASCO FARIS-1 spectrometer. 

\begin{figure}[bht]
\centerline{\includegraphics[width=8.5cm,clip=true]{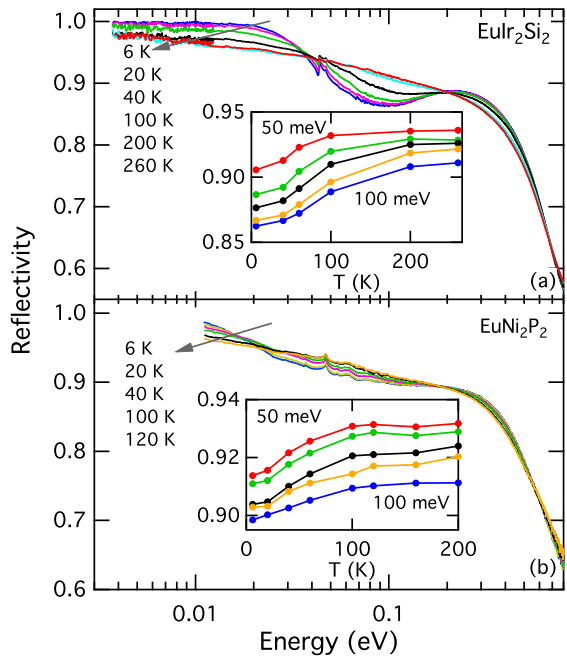}}
 \caption{(Color online) 
Reflectivity spectra of  EuIr$_2$Si$_2$ and EuNi$_2$P$_2$ at various temperatures.  For EuIr$_2$Si$_2$ the sharp peak at 44\,meV is an optical active phonon mode. The structure seen at 47\,meV in the spectra of both materials is an experimental artifact. Insets: The temperature dependence of $R(\omega)$ at 50, 60, 68, 80 and 100\,meV (from top to bottom).} 
\label{r} 
\end{figure}

Figures \ref {r}(a) and \ref {r}(b) display the infrared reflectivity spectra $R(\omega)$ of EuIr$_2$Si$_2$ and EuNi$_2$P$_2$ single crystals for various temperatures down to 6 K. The close resemblance in general shape and trends of changes for both materials indicates that their electrodynamic properties are qualitatively similar. At low energy we observe that $R(\omega)$ increases with decreasing temperature and approaches unity when $\omega\rightarrow0$ as expected for a metal. On the other hand, in the energy range between 30 meV and 200 meV $R(\omega)$ is strongly suppressed, and a dip structure develops upon cooling, which is more pronounced for EuIr$_2$Si$_2$. This suppression in the reflectivity spectra at low temperatures signals the development of a pseudogap. Interestingly, the temperature dependence of the spectra shown in the insets in Fig. \ref {r} becomes stronger below 200\,K (100\,K) for EuIr$_2$Si$_2$ (EuNi$_2$P$_2$), where the resistivity shows a sharp decrease. Besides these features, for EuIr$_2$Si$_2$ a sharp peak at 44 meV emerges in the spectra  due to an infrared active phonon. 

\begin{figure}[bht]
\centerline{\includegraphics[width=8.5cm,clip=true]{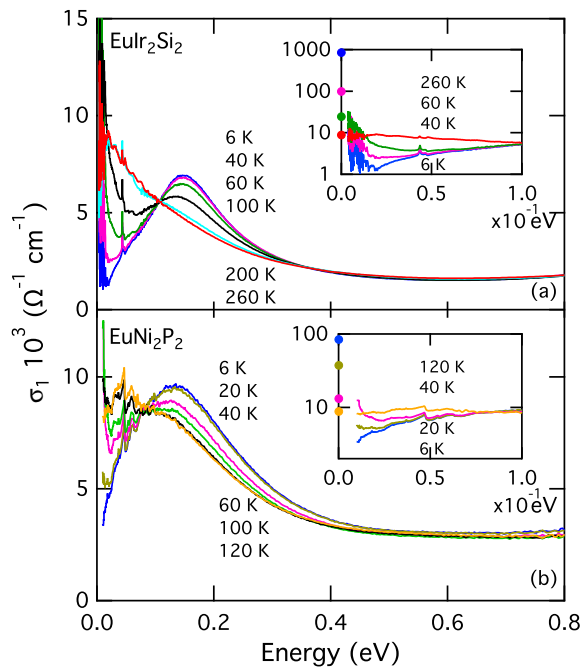}}
 \caption{(Color online) 
The real part of the optical conductivity  spectra of EuIr$_2$Si$_2$ and EuNi$_2$P$_2$ at various temperatures. Insets: The low-energy optical conductivity spectra (solid lines) and the corresponding values for the dc conductivity, $\sigma_{\rm dc}$\,=\,$\rho^{-1}$, (symbols) at 6 , 40, 60 and 260\,K for EuIr$_2$Si$_2$ and at 6, 20, 40 and 120\,K for EuNi$_2$P$_2$.}
\label{s1} 
\end{figure}

The  optical conductivity was derived from the measured reflectivity spectra in combination with the independently measured dc conductivity data using a Kramers-Kronig consistent variational fitting procedure  \cite{kuzmenko2005}. Figures \ref {s1}(a) and \ref {s1}(b) show the real part of the optical conductivity spectra $\sigma_{1}(\omega$) of EuIr$_2$Si$_2$ and EuNi$_2$P$_2$. As the temperature is lowered three remarkable structures emerge in the optical conductivity of both materials. First, at the lowest measured energy the tail of a narrow Drude-like peak is observed, ascribed to the intraband response of heavy quasiparticles. The data reveal that the dc conductivity ratio $\sigma_{1}(\omega$ = 0, 6\,K)/$\sigma_{1}(\omega$ = 0, 260\,K) increases by a factor of 100 for EuIr$_2$Si$_2$ and $\sigma_{1}(\omega$ = 0, 6\,K)/$\sigma_{1}(\omega$ = 0, 120\,K) $\sim 10$ for EuNi$_2$P$_2$ as shown in the insets from  Fig. \ref {s1}. This implies the appearance of a very narrow Drude-like peak in the low-energy optical conductivity of both materials. Second, the optical conductivity strongly decreases below 100\,meV with decreasing temperature. This leads to a depletion of the spectral weight immediately above the narrow Drude-like peak indicating the opening of a pseudogap. The third feature is the MIR peak centered at 0.15\,eV for EuIr$_2$Si$_2$ and at 0.13\,eV for EuNi$_2$P$_2$. Note that this is the first experiment which reveals the appearance of a MIR peak in VF Eu-based compounds. The peak maximum is slightly shifted toward lower energy with increasing temperature and vanishes above $\approx100$\,K. 

In VF systems, it seems natural to relate the MIR peak to the energy difference between the divalent and the trivalent configurations. However, using the phenomenological model of interconfiguration fluctuations \cite{sales1975} to fit the susceptibility $\chi(T)$ or the valence $\nu(T)$ results in an energy difference of about 200\,K which is one order of magnitude smaller than the energy of the MIR peak ($\equiv1600$~K). In the Falicov-Kimball (FK) model \cite{zlatic2001}, being the basic Hamiltonian describing the VF systems, the distinctive ingredient is the on-site Coulomb repulsion $U_{fc}$ between an $f$ electron and a conduction electron. This model predicts a MIR peak located at U$_{fc}$, while the valence transition occurs at a temperature much lower in comparison to U$_{fc}$. The energy of the MIR peak is an order of magnitude larger than the temperature of the maxima in $\rho(T)$ (see Fig.~\ref{rho_valence}) or $\chi(T)$, in qualitative agreement with the FK model. However, this model predicts a decrease of the intensity of the MIR peak with decreasing temperature, while the experiment shows an increase of its intensity upon cooling, in strong disagreement with the calculations.  As suggested in Ref. \cite{zlatic2001} this might be related to the absence of any hybridization term in the FK model.

\begin{figure}[bht]
\centerline{\includegraphics[width=10cm,clip=true]{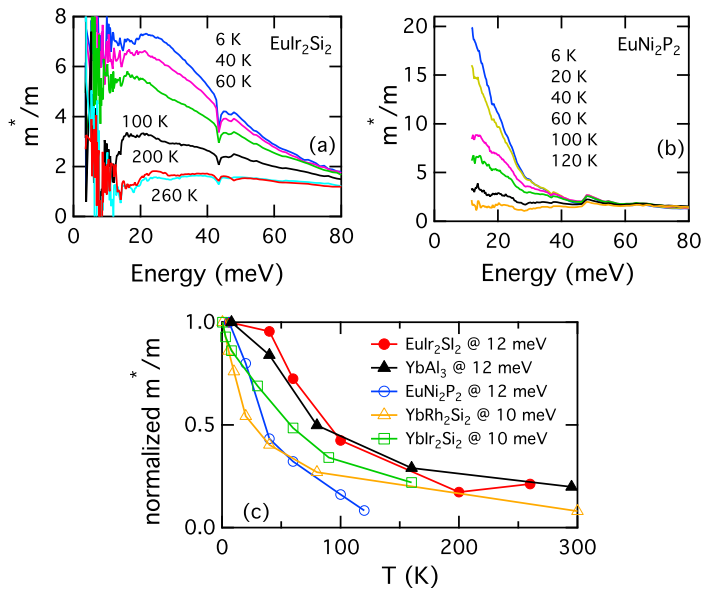}}
 \caption{(Color online) The effective mass relative to the band mass as a function of energy for EuIr$_2$Si$_2$ and EuNi$_2$P$_2$. The normalized effective mass (c) for  EuIr$_2$Si$_2$ and EuNi$_2$P$_2$, as a comparison, the data for YbIr$_2$Si$_2$ \cite{iizuka2010}, YbRh$_2$Si$_2$  \cite{kimura2006} and YbAl$_3$ \cite{okamura2004} are shown.} 
\label{mstar} 
\end{figure}

The relevance of hybridization is indicated by the clear similarity of the main features in our optical spectra with those reported for Ce- or Yb-based KL/IV systems \cite{garner2000,mena2005,okamura2004,degiorgi2001}. There, features like the narrow Drude-like peak, the pseudogap, or the MIR peak are discussed in terms of the hybridization between localized $f$ electrons and conduction electrons. This observation is quite remarkable since the characteristic phase diagram for the VF Eu-systems as a function of $x$ or $p$ as well as the temperature evolution of the valence are very different from those of Ce- or Yb-based  KL/IV systems \cite{seiro2011}. Therefore, one would expect the optical spectra and their energy and temperature dependence to be very different. Thus, the optical data for EuIr$_2$Si$_2$ and EuNi$_2$P$_2$ strongly indicate that the hybridization between 4$f$ and conduction electrons play an important role in understanding the low-energy optical spectra of VF Eu-compounds. These observations are supported by recent angle-resolved photoemission measurements on EuNi$_2$P$_2$ where localized 4$f$ states were observed to hybridize with valence states as in Ce- or Yb-based KL systems, resulting in 4$f$ dispersion and the opening of hybridization gaps \cite{danzenbacher2009}.\\
In order to further analyze the temperature and energy evolution of the low-energy excitations we plot in Figs. \ref {mstar}(a) and \ref {mstar}(b) the energy dependence of the effective mass $m^\ast(\omega)$ relative to the band mass $m$ obtained from the extended Drude analysis \cite{puchkov1996}: $\frac{m^\ast(\omega)}{m}$\,=\,$\frac{\omega_{\rm p}^{2}}{4\pi\omega}$Im$\left[\frac{1}{\sigma(\omega)} \right]$. Here, one should note that this model is meaningful in the energy range below the interband transitions, where the optical conductivity is dominated by the response of mobile carriers. We assume that at energies lower than 100\,meV the contribution of interband transitions to the optical conductivity is negligible and this model can be applied. Therefore, we determined the plasma frequencies $\omega_{\rm p}$\,=\,3\,eV (3.9\,eV) through $m^{\ast}/m$\,=\,1 at $\omega$\,=\,80\,meV for EuIr$_2$Si$_2$ (EuNi$_2$P$_2$). It should be mentioned that the magnitude of $m^{\ast}/m$ depends on the chosen values for the plasma frequency, but the general shape is unaffected by the choice of $\omega_{\rm p}$. For both materials we observe at low temperature a strong energy and temperature dependent effective mass indicating deviations from the simple Drude response. The data reveal that the energy and temperature scales are larger for EuIr$_2$Si$_2$. At 6\,K and the lowest measured energy we find $m^{\ast}/m$\,$\sim$\,8 (20) for EuIr$_2$Si$_2$ (EuNi$_2$P$_2$). These values are in good agreement with results of specific heat measurements, where the ratio of the Sommerfeld coefficient of EuIr$_2$Si$_2$ (EuNi$_2$P$_2$) to that of its non-correlated homologue LaIr$_2$Si$_2$ (LaNi$_2$P$_2$) amounts to 7.3 (17.5) \cite{seiro2011,fisher1995}. Such a moderate mass enhancement is comparable to values reported for  Ce- or Yb-based IV compounds with a similar characteristic 4$f$ energy.\\
Notwithstanding the optical similarities to the KL/IV systems specific differences regarding the effective mass of VF Eu-systems can be identified. In Fig. \ref {mstar}(c) we compare the temperature evolution of the effective masses normalized to their low-temperature values for EuIr$_2$Si$_2$ and EuNi$_2$P$_2$ with the data for the heavy-fermion metals YbIr$_2$Si$_2$ \cite{iizuka2010} and YbRh$_2$Si$_2$ \cite{kimura2006} as well as the IV compound YbAl$_3$ \cite{okamura2004}. These Yb-systems share the following features in common with the EuIr$_2$Si$_2$ and EuNi$_2$P$_2$: the narrow Drude peak, the pseudogap formation and the MIR peak. Although the general tendency in the temperature dependence of the normalized $m^{\ast}/m$ is similar for all materials, a closer look reveals characteristic differences, which can be linked to the different nature of each compound.  For YbRh$_2$Si$_2$, YbIr$_2$Si$_2$, and EuNi$_2$P$_2$ the effective mass drops rapidly upon increasing temperature. In contrast, in the archetypical VF Eu-system EuIr$_2$Si$_2$, the effective mass remains almost constant up to 40\,K where a steep drop sets in. The low-temperature behavior in the IV system YbAl$_3$ is in-between, with some evidence for a negative curvature around 40\,K. 

A further difference can already be discerned in Fig. \ref {mstar}(c), but becomes even more pronounced when the absolute values of $m^{\ast}/m$ are considered. The mass renormalization around 100\,K is still quite significant in YbRh$_2$Si$_2$ and YbIr$_2$Si$_2$ ($m^{\ast}/m$\,$\sim$\,50) \cite{kimura2006,iizuka2010} and in YbAl$_3$($m^{\ast}/m$\,$\sim$\,15) \cite{okamura2004} while in EuIr$_2$Si$_2$ and EuNi$_2$P$_2$ no significant enhancement is observed ($m^{\ast}/m$\,$\sim$\,3). This is in line with expectations from the underlying models. The Anderson model \cite{hewson1993} being based on hybridization effects with conduction electrons results in a logarithmic temperature dependence at high temperatures, which implies some renormalization up to very high temperatures. In contrast, for VF systems close to the first-order valence transition like EuIr$_2$Si$_2$, the FK model predicts strong changes in a narrow temperature range. However, the differences between our results on the VF Eu-systems and those reported for Ce- or Yb-based KL/IV systems are much less pronounced than one might have expected from these models.

In conclusion, we have studied the optical properties in the infrared energy range of the prominent VF systems EuIr$_2$Si$_2$ and EuNi$_2$P$_2$.  For both materials we observe upon decreasing temperature the formation of a narrow Drude-like response, a partial suppression of the optical conductivity below 100\,meV and a MIR peak at 0.15\,eV for EuIr$_2$Si$_2$ and at 0.13\,eV for EuNi$_2$P$_2$.  These results present striking similarities to those reported for Ce- or Yb-based KL/IV systems, despite the much stronger change of the valence as a function of temperature and a completely different phase diagram as a function of pressure or chemical tuning in VF Eu-systems. This highlights the importance of hybridization between $f$- and conduction electrons in the VF Eu-systems and calls for further theoretical calculations of the optical properties for a Hamiltonian combining a hybridization term with an intra-site Coulomb repulsion between $f$- and conduction electrons. An examination of the energy dependence of the effective mass derived from the optical data revealed specific differences regarding the temperature evolution of the effective mass. Our results are in line with recent developments challenging the KL/VF dichotomy, like the second superconductivity dome found at higher pressure in the paradigmatic heavy fermion system CeCu$_2$Si$_2$ which is proposed  to be mediated by critical valence fluctuations \cite{yuan2003, holmes2004}. 

We thank Dr. A. Irizawa for kindly sharing his beam time at the UVSOR. We also acknowledge S. Kostmann and K. Imura for technical assistance and H. Okamura for sharing his optical data for YbAl$_3$. V. G. benefited from financial support from the Alexander von Humboldt Foundation.

\end{document}